%% file: main.tex
\documentclass[10pt,letterpaper]{article}

\pdfoutput = 1

\usepackage{ccn}
\usepackage{pslatex}
\usepackage{apacite}
\usepackage{hyperref}
\usepackage{natbib}
\usepackage{lineno}
\usepackage{graphicx}
\usepackage{amsmath,amssymb,amsfonts}
\usepackage{xcolor}
\usepackage{bm}

\usepackage{tikz}
\usetikzlibrary{bayesnet}
\usetikzlibrary{calc}

\title{A flexible Bayesian non-parametric mixture model reveals multiple dependencies of swap errors in visual working memory}
 
\author{
  \textbf{Puria Radmard}\\
  Department of Engineering\\
  University of Cambridge\\
  \texttt{pr450@cam.ac.uk}
  \AND
  \textbf{Paul M. Bays}\\
  Department of Psychology\\
  University of Cambridge\\
  \texttt{pmb20@cam.ac.uk}
  \AND
  \textbf{Máté Lengyel}\\
  \begin{minipage}[t]{0.4\textwidth}
    \raggedleft
    Department of Engineering\\
    University of Cambridge
  \end{minipage}
  \hspace{0.5cm}
  \begin{minipage}[t]{0.4\textwidth}
    \raggedright
    Department of Cognitive Science\\
    Central European University
  \end{minipage}\\[5pt]
  \texttt{m.lengyel@eng.cam.ac.uk}
}

\begin{document}

\maketitle

\section{Abstract}
{
\bf
\input{abstract}
}
\begin{quote}
\small
\textbf{Keywords:} 
visual working memory; psychophysics; binding; short-term memory; swap error; feature binding
\end{quote}

\input{graphical_model}

\section{Introduction}
\input{1_introduction}
\input{figures/CCN_figure1}

\input{figures/CCN_figure2}

\section{Model description and validation}
\input{2_model_description}

\input{figures/CCN_figure3}

\subsubsection{Model validation}
\input{3_model_validation}

\input{figures/CCN_figure4}

\input{figures/CCN_figure5}

\section{Results}
\input{4a_datasets}
\subsubsection{Recapturing probe distance dependence}
\input{4b_simple_cue_recovery}
\subsubsection{Non-trivial probe dependencies} 
\input{4c_non_trivialcue_recovery}

\input{figures/CCN_figure6}

\subsubsection{Mixed dependence}
\input{5a_rdk_mixed_dependence}

\subsubsection{Results validation and model recovery} 
\input{5b_rdk_mixed_dependence_validation}

\section{Discussion}
\subsubsection{Significance} \input{6a_significance}

\subsubsection{Limitations and future work} \label{sec:limitations} \input{6b_limitations}



\bibliographystyle{ccn_style}



\end{document}

%% file: abstract.tex
Human behavioural data in psychophysics has been used to elucidate the underlying mechanisms of many cognitive processes, such as attention, sensorimotor integration, and perceptual decision making. Visual working memory (VWM) has particularly benefited from this approach: analyses of VWM errors have proven crucial for understanding VWM capacity and coding schemes, in turn constraining neural models of both. One poorly understood class of VWM errors are swap errors, whereby participants recall an uncued item from memory. Swap errors could arise from erroneous memory encoding, noisy storage, or errors at retrieval time - previous research has mostly implicated the latter two. However, these studies made strong a priori assumptions on the detailed mechanisms and/or parametric form of errors contributed by these sources. Here, we pursue a data-driven approach instead, introducing a Bayesian non-parametric mixture model of swap errors (BNS) which provides a flexible descriptive model of swapping behaviour, such that swaps are allowed to depend on both the probed and reported features of every stimulus item. We fit BNS to the trial-by-trial behaviour of human participants and show that it recapitulates the strong dependence of swaps on cue similarity in multiple datasets. Critically, BNS reveals that this dependence coexists with a non-monotonic modulation in the report feature dimension for a random dot motion direction-cued, location-reported dataset. The form of the modulation inferred by BNS opens new questions about the importance of memory encoding in causing swap errors in VWM, a distinct source to the previously suggested binding and cueing errors. Our analyses, combining qualitative comparisons of the highly interpretable BNS parameter structure with rigorous quantitative model comparison and recovery methods, show that previous interpretations of swap errors may have been incomplete. These findings necessitate further systematic investigation, and demonstrate the viability of our data-driven Bayesian approach for driving future experimental design.

%% file: graphical_model.tex
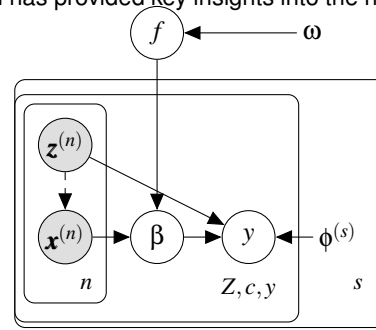
\begin{figure}[b!]
\label{fig:graphical_model}
	\centering
    \vspace{-2.5em}
\begin{tikzpicture}


    \node[obs]                (Z)               {$\boldsymbol{z}^{(n)}$} ; %
    \node[obs, below=0.5cm of Z]    (X)               {$\boldsymbol{x}^{(n)}$} ; %
    \node[latent, right=0.5cm of X]    (beta)         {$\beta$} ; %
    \node[latent, above=2.0cm of beta]    (f)         {$f$} ; %
    \node[latent, right=0.5cm of beta]    (y)     {$y$} ; %
    \node[const, right=0.5cm of y]    (phi)   {$\phi^{(s)}$} ; %
    \node[const, right=1.5cm of f]    (omega)           {$\omega$} ; %



    \edge[dashed]{Z}{X}
    \edge[]{X, f}{beta}
    \edge[]{beta}{y}
    \edge[]{phi, Z}{y}
    \edge[]{omega}{f}

    \plate [inner sep=.3cm] {data} { %
        (Z)(X)
        (beta)(y)
    } {$Z, c, y$}; %

    \plate[inner sep=.3cm, inner ysep=.5cm] {subject} { %
        (Z)(X)
        (beta)(y)
        (phi)
    } {$s$}; %

    \plate {stimulus} { %
        (Z)(X)
    } {$n$}; %

\end{tikzpicture}

	\vspace{4pt}
    \caption{Graphical model of BNS, described in detail in the main text}
\end{figure}

%% file: 1_introduction.tex
Delayed estimation is a dominant task paradigm in the study of visual working memory (VWM).
Compared to change detection \citep{Luck1997, Keshvari2013}, the continuous nature of the response provides a much richer description of the error in VWM.
Of these, a frequently used task structure is multiitem delayed estimation, where the subject is presented with an an array of multiple distinct visual stimuli which vary along at least two feature dimensions (location, orientation, colour, etc.).
After a short delay, one feature of one item is again presented, and the subject is asked to reproduce the other feature of the corresponding item in memory.
In the trial-by-trial response data, the frequency of responses typically peaks near the correct report dimension feature value, and naturally the precision of this peak decreases as the number of items - the \textit{set size} - typically increases.
The form of this degradation has provided key insights into the nature of VWM and its interactions with other cognitive processes \citep{Ma2014, Schneegans2016}.

However, this degradation in precision is accompanied by another, less well understood response pattern in multiitem delayed estimation, where subjects mistakenly reproduce the report feature value of an \textit{uncued} item.
In response data, these \textit{swap errors} are observed as a central tendency of responses towards non-target feature values \citep{Bays2016}.
This has motivated the \textit{three-component mixture model} where trial-by-trial continuous working memory response data is decomposed into 3 sources: a component centered at the target feature value representing accurate albeit noisy recall; a uniform component over the response feature dimension, represent random guesses; and component(s) centered at the non-target feature values, representing swap errors \citep{Bays2009, Bays2011, Gorgoraptis2011}.
A variant that places a probabilistic prior over the weight of this component is used for modeling swap errors in monkey behavior datasets \citep{Alleman2024}.
While maximum likelihood is used to fit the parameters of these mixture models, \cite{Bays2016} offers a non-parametric alternative which removes the uniform component and makes no \textit{a prior} assumptions of the shape of correct and swap components.

There are multiple hypothesised neural mechanisms for swap errors.
They can arise during encoding at stimulus presentation time, with items written to memory with incorrect associations between cue and response features.
Correctly encoded bindings could also become corrupted in memory, leading to \textit{misbinding} \citep{Swan2014}.
Finally, swap errors could arise at cuing time, with noise in the probe feature dimension causing the incorrect, but correctly bound, item to be retrieved from memory.
Past studies have largely pointed to the latter, given that swap errors strongly correlate with proximity in the cue feature dimension, and that the probability of swap errors depends on which feature is being cued \citep{Emrich2012, Oberauer2017, McMaster2022}.
Neural models of swap errors therefore often involve a failure during the decoding of a target feature from a stored representation of the full array of items \citep{Matthey2015, Schneegans2017}.

Here, we revisit the data used to constrain these neural models, and reassess the relationship between feature similarity and swap errors.
We design a \textbf{B}ayesian \textbf{n}on-parametric model of \textbf{s}wap errors (BNS) that allows the \textit{a priori} probability of swapping to each uncued item to depend fully on its distance to the cued item in \textit{both} cue \textit{and} report dimensions.
Importantly, we include dependence on uncued features in the generative process of our model, unlike previous approaches which only allow post hoc correlatory analysis \citep{Bays2016}, while also making no assumptions about the form this dependence, unlike existing neural models of swap errors.
We find evidence that distance in the report dimension can play a role in swap errors, implying a failure at stimulus presentation time leading to a failure to encode bindings correctly, challenging the dominance of retrieval-based explanations in current literature.
We verify our findings using rigorous Bayesian model comparison and recovery techniques, and demonstrate that our assumption-free approach reveals complex dependency structures that would remain hidden with more constrained models.
We propose our data-driven Bayesian framework as a powerful tool for driving both future psychophysics experiments and neural modeling, offering a path to more comprehensive understanding of the multifaceted mechanisms underlying visual working memory failures.

%% file: figures/CCN_figure1.tex
\begin{figure*}[!ht]
\begin{center}
  \includegraphics[width=0.8\linewidth]{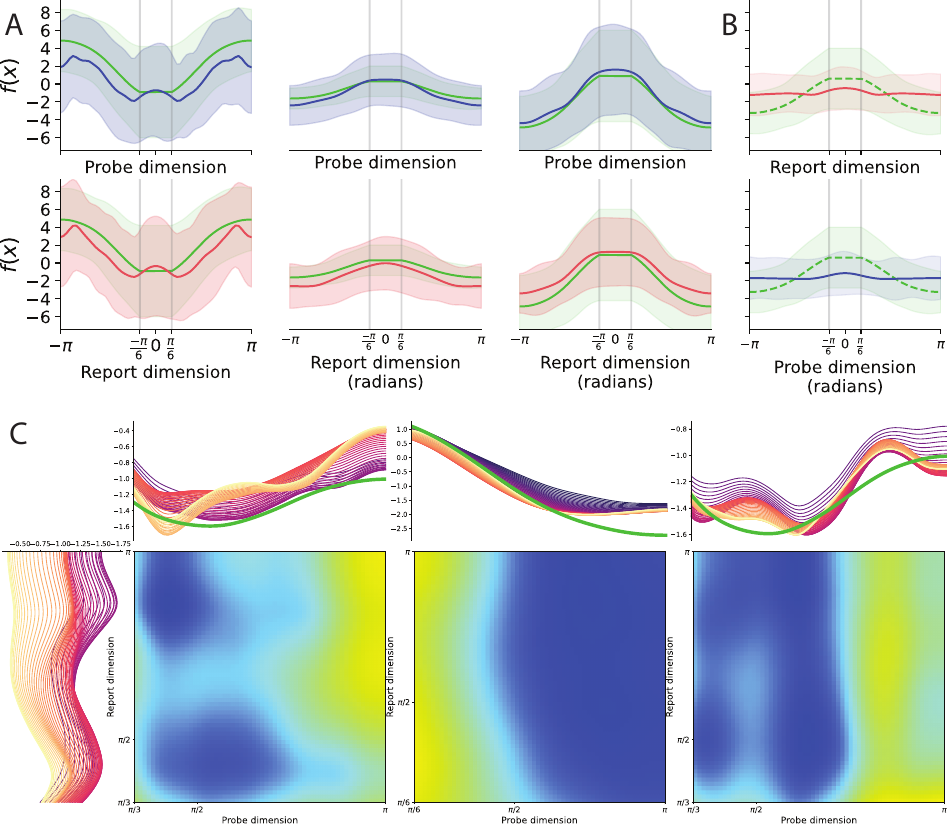}
\end{center}
\caption{
    Variational approximations of the swap function posterior given synthetic data $q(f)\approx p(f|\mathcal{D}[\theta_\mathcal{M}])$ -
    \textbf{A}: when fit by the same model structure as the generative model. 
    The distribution $p_f(f)$ used to generate the data is given in green, and is augmented by scaling to show the BNS's versatility in recovering different true underlying swap functions. Note that $f(\boldsymbol{x}) > 1$ is unlikely in real data, as this implies distractors displaced by $\boldsymbol{x}$ in feature space are more likely to be recalled than the cued item.
    \textbf{B}: when the fitted model has a different structure $\mathcal{M}'\neq\mathcal{M}$ to the generative model. The green posterior is shown as dotted as it does not span the same axes as the infered blue/red posterior, instead spanning the probe (report) axes in the top (bottom) axes.
    \textbf{C}: when the fitted model has structure $\mathcal{M}'=\texttt{both}$, and the generative model $\mathcal{M}=\texttt{probe}$. 
    Heatmaps show the swap function mean under $q(f)$ in the positive quadrant (excluding the miniumum margin of seperation from the cued item), and inset axes show slices, with darker slices being closer to the cued item in feature space.
    Bright green indicates large mean of $f$, and dark blue low.
    In all three cases, the true generative distribution $p(f)$ (of which only the mean is shown) is learned from data (see Figure \ref{fig:visualise_real_data_fits} and main text). Left and center: BNS is fitted with a realistic number of trials, and is able to remove \textit{most} of the variability in the redundant dimension. Right: the same ground truth as in the leftmost plot is refitted with a much larger (10x) amount of data.
    For ease of viewing, variance of the swap functions are not displayed, nor are the exponentiated swap functions (relevant due to the softmax step in the generative process).
    BNS's accuracy in capturing the shape of the underlying swap function wavers mostly at unrealistically high values (\textbf{A}, left), or extremely low values which are compressed during exponentiation (\textbf{C}, middle).
} 
\label{fig:model_validation}
\end{figure*}

%% file: figures/CCN_figure2.tex
\begin{figure*}[t!]
\begin{center}
  \includegraphics[width=\linewidth]{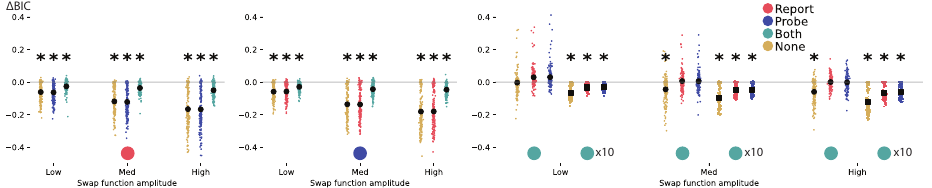}
\end{center}
\caption{
$\Delta\text{BIC}_s(\mathcal{M};\mathcal{M}')$ (average per-trial difference; see main text) for all combinations of model classes. The swap function amplitude refers to modulation of swap dependence in the generating model parameters. Some examples of different amplitudes are shown in Figure \ref{fig:model_validation}A.
    The data generating model $\mathcal{M}$ is denoted by the dots on the abscissa.
    In most cases, each model is fit with data from 10 `synthetic subjects', each performing 96 trials (see \textbf{Materials}), and scattered coloured dots indicate individual synthetic subjects' $\Delta\text{BIC}_s$ values for that model combination, across many such synthetic datasets and fitted models.
    Asterisks denote these values fall significantly ($p<0.05$) below 0, indicating successful model recovery for that combination.
    Where marked, synthetic datasets with 10 synthetic subjects performing 960 tasks were generated and fitted for each synthetic dataset.
    Black marks represent overall mean for that model class combination.
} 
\label{fig:model_validation_BIC}
\end{figure*}

%% file: 2_model_description.tex
\subsubsection{Task notation}
We consider simple multiitem VWM tasks, where the human subject is presented with a stimulus array of $N$ items, composed of a circular $p$robe and $r$eport-dimension feature values: $Z = \left\{\bm{z}^{(n)} = [z_p^{(n)}, z_r^{(n)}]^\intercal: n=1,...,N \right\}$.
Circular features $z_p$, $z_r\in[-\pi,+\pi)$ in datasets we consider are location on an (invisible) circle, colour on a colour wheel, random dot kinematogram (RDK) movement direction, ellipse/grating orientation (on a half circle; extended to a $2\pi$ period), etc.
After stimulus exposure, there is a short delay, followed by presentation of the cue, equal to the probe feature value $z_p^{(1)}$ of item $n=1$ - by convention we always take this item to be the cued one.
The subject then generates an estimate $y\in[-\pi,+\pi)$ - their recollection of $z_r^{(1)}$.
Each dataset is composed of $M$ stimulus array-subject-estimate tuples: $\mathcal{D} = \{Z^{(m)}, y^{(m)}, s^{(m)}\}_{m=1}^M$, where $s^{(m)}\in[S]$ is an index denoting which of the $S$ subjects performed trial $m$.
In each dataset, subject $s$ performs $M_s$ trials.


\subsubsection{Generative process}

The BNS models the estimate $y$ as being drawn from a mixture distribution defined on a circle, doubly conditioned on the stimulus array $Z$.
This dependence is depicted in Figure 1, and is presently described.
First, mixture components are centered on each item $z_r^{n}$, with an optional uniform component, as in previous mixture models of swap errors \citep{Bays2009}, 
Second, the component weights depend on the stimulus array via a \textit{swap function} $f$, which maps elements of $Z$ (stimulus feature values) to logits (unnormalised log-probabilties).
%
%
Below we present the full generative model $p(y | s, Z)$, after which we will dissect each stage:
\begin{align*}
    \bm{x}^{(n)} &= \bm{z}^{(n)} \ominus \bm{z}^{(1)}, n = 2, ..., N \\
    f &\sim p_f(f) \\
    \tilde{\bm\pi} &= [\tilde\pi_0, \tilde\pi_1, f(\bm{x}^{(2)}), ..., f(\bm{x}^{(N)})] \quad\quad
    \bm\pi = \text{Softmax}(\tilde{\bm\pi}) \\
    \beta &\sim \text{Cat}(\bm\pi) \\ 
    y&=\begin{cases}
        \varepsilon\oplus z_r^{(\beta)} \quad\quad\quad\quad \varepsilon \sim p_\varepsilon(\cdot;\phi^{(s)}) & \beta > 0\\
        \varepsilon_u\sim\mathcal{U}[\pi,+\pi)   & \beta = 0
    \end{cases}
\end{align*}
The vectoral displacement $\boldsymbol{x}^{(n)}$ of each uncued stimulus ($n > 1$; hereafter: distractor) from the cued item is found with elementwise circular subtraction.
The swap function $f$ is drawn from a Gaussian Process (GP) prior $p_f(f)$ over $f$ \citep{gps} with kernel $k(\cdot,\cdot';\omega_k)$ described below, and evaluated at each distractor's displacement.
Alongside $\tilde\pi_0$ and $\tilde\pi_1$, these form the logits of the mixture distribution component weights, $\boldsymbol{\pi}$.
The estimate type is then drawn from this mixture distribution.
An estimate drawn from component $\beta=0$ indicates a uniform guess over the circle.
An estimate drawn from $\beta=1$ indicates a correct choice is made, and $\beta > 1$ indicates a swap error to item $n=\beta$.
In these latter two cases, a \textit{residual} is drawn from some zero-mean emission distribution $p_\epsilon$ over the circle, and the estimate is the circular addition of the relevant report feature value and this residual.
The emission distribution is shared by all non-uniform components.
$\tilde\pi_0$ is either set to $-\infty$ (no uniform component included in the final distribution), or is learned alongside the parameters of the GP kernel $\omega_k$, collectively forming the \textit{swap function generative parameters} $\omega$.
These parameters are shared for all $s$ - this is a model limitation that is addressed in at the end of this work.
We fix $\tilde\pi_1$, the logit for a correct choice, to $1$.
Emission distribution parameters $\phi^{(s)}$ are subject-specific.
\\

\subsubsection{Swap function and GP prior}

The swap function takes in indiviaul distractor displacements, and provides logits for our mixture model over swap errors.
Although we always denote $f$ as taking a vector input, it can be chosen to depend on either probe or report dimensions, both of them, or neither.
We denote this model choice as $\mathcal{M}=\texttt{probe}, \texttt{report}, \texttt{both}, \texttt{none}$ respectively.
The $\texttt{none}$ case is equivalent to placing a scalar Gaussian prior over a shared swap logit: $f(\boldsymbol{x})=\tilde\pi_s\sim\mathcal{N}(\mu_\omega,\sigma_\omega^2)$, i.e. $\omega_k=\{\mu_\omega,\sigma_\omega^2\}$.
This is roughly equivalent to mixture model used by \citet{Alleman2024}, and after marginalisation over $\tilde{\pi}_s$, is equivalent to the simple three-component of \citet{Bays2009}.
For $\texttt{probe}$, we instead place a zero mean GP prior over $f$ - $f \sim \mathcal{GP}(0, k(\cdot,\cdot';\omega_k))$, with:
\begin{align}
    k(\bm{x},\bm{x}';\omega_k) &= 
        \sigma^2_0 \delta(x_p - x_p')+ \sigma^2_\omega W(x_p-x_p';\tau) 
    \\
    W(d;\tau) &= (1 + \tau\frac{d}{2\pi}) [1-\frac{d}{2\pi}]^\tau_+
\end{align}
and $\omega_k = \{\sigma^2_\omega, \tau\}$, where the Weinland kernel function $W$ could also act on $x_r,x_r'$, as in the $\texttt{report}$ model.
The two-dimensional case, $\texttt{both}$, has a covariance kernel of the form:
$k(\bm{x},\bm{x}';\omega_k) = 
        \sigma^2_0 \delta(\bm{x} - \bm{x}')
        + \sigma^2_\omega W(x_p-x_p';\tau_p) W(x_r-x_r';\tau_r)$, 
        requiring parameters   
    $\omega_k = \{\sigma^2_\omega, \tau_p, \tau_r\}$

\subsubsection{Relation to mechanisms underlying swap errors}

The $\texttt{probe}$ model architecture conditions the probability of swap errors purely on proximity in the probe feature dimension. Such a dependence indicates a failure of memorandum retrieval at cuing time, i.e. after the delay in the experimental paradigm we considered. \citet{Schneegans2017} build neural population models to predict swap errors due to this retrieval failure, but $f$ learns this dependence without assumptions about its parametric form.
The $\texttt{report}$ model architecture conditions the probability of swap errors purely on proximity in the same feature dimension in which the subject makes an estimate. For a location-report experiment, poor encoding would lead to a misbinding of probe features to precisely represented \citep{McMaster2022} locations, causing swap errors to spatially proximal memoranda. Again, BNS can learn the form of this dependence on spatial proximity in a principled way. We will shortly reveal a report-dimension dependence in such a dataset.
Note that while an encoding error cannot manifest as a dependence in the probe dimension (the cued feature is unknown at the time of stimulus encoding), we have not ruled out the possibility of a report dimension dependence also arising from retrieval error. Regardless, BNS’s ability to capture both of these dependencies, both independently and concurrently (in the $\texttt{both}$ model), provides a key advantage over prior models which required assumptions about both the cognitive processes underlying swap errors, and their parametric forms.

\subsubsection{Emission distribution}
After a component is selected, a residual is drawn from $p_\epsilon(\cdot; \phi^{(s)})$.
We consider only models that have: (i) a von Mises emissions distribution with a uniform guess component, or; (ii) an (unskewed) wrapped stable \citep{Pewsey2008,vandenBerg2012} with no uniform component.
In each case there are: (i) one emission parameter per subject (concentration $\kappa$) and one shared swap model generative parameter (uniform scalar $\tilde\pi_0$), and; (ii) two emission parameters per subject (stability parameter $\alpha$ and scale parameter $\gamma$).
The form of this emissions distribution is not central to most of our investigation, and unless stated otherwise results are given with a von Mises emission distribution.


\subsubsection{Sparse variational inference of swap function posterior}

Gaussian processes enjoy a closed form posterior $p(f | \mathcal{D}) \propto p(f) p(\mathcal{D} | f)$ in the event that the downstream likelihood function of the data given a draw from the GP prior is Gaussian and independent between inputs.
Unfortunately, this is not the case for our model: our likelihood function has a hierarchical structure, and contains interdependence on function evaluations via the softmax function.
The over likelihood function downstream of swap function $f$ is:
\begin{align}
\label{eq:actual_likelihood}
    p(y | f, s, Z) = \frac{ \frac{1}{2\pi}e^{\tilde\pi_0} + \sum_{n=1}^N e^{\tilde\pi_n} p_\epsilon(y - z_r^{(n)}; \psi^{(s)})}{\sum_{n=0}^N e^{e^{\tilde\pi_n}}}
\end{align}
with elements of $\tilde{\bm\pi}$ defined based on $f$ previously.
To approximate this intractable posterior, we use a sparse variational GP (SVGP) approximation $q(f;\psi)$, with \textit{variational parameters} $\psi$.
The parameters $\theta = \{\omega, \psi, \phi\}$ are jointly trained on the joint objective of \textit{maximising marginal (log)likelihood over the data} while \textit{minimising the KL divergence between $q(f)$ and the true posterior}.
\citet{svgp} show that this is equivalent to maximising an evidence lower bound (ELBO) with no access to the true posterior.
The exact form of our variational approximation $q(f)$ is omitted; we constrain the mean function of this approximation to be symmetric around $\boldsymbol{x}=0$, and in many cases, we will only display the inferred $q(f)$ in the positive quadrant.
For the $\texttt{none}$ model, there is no swap function, so the variational approximation is just a Gaussian over $\tilde\pi_s$.

\subsubsection{ELBO and BIC approximation}

To evaluate model performance in fitting data from subject $s$, we disaggregate the average Bayesian information criteria (BIC) across all $M$ trials with $M\cdot\text{BIC}(\mathcal{D}; \theta) = \sum_sM_s \cdot\text{BIC}_s(\mathcal{D}_s; \theta)$, yielding subject-specific average per-trial BIC:
\begin{equation}
\begin{split}
    \text{BIC}_s(\mathcal{D}_s; \theta) = -&\frac{1}{2M}|\omega|\log M - \frac{1}{2M_s}|\phi^{(s)}|\log M \\ + &\frac{1}{M_s} \sum_{s^{(m)}=s} \log p(y^{(m)} | s^{(m)}, Z^{(m)}) 
\end{split}
\end{equation}
Note that this disaggregation of BIC is largely heuristic --defining a subject-specific marginal log-likelihood is challenging since subject-specific datasets aren't fitted independently -- but allows pairwise comparisons of subjects across models.
We approximate the marginal likelihood (with respect to $f$) with: $p(y | s, Z) \approx \sum_{\beta} p(y | \beta, s, Z) \sum_k p(\beta|f^{(k)}, Z)$, with $f^{(k)} \overset{i.i.d.}{\sim} q(f)$.
The approximation arises both because we are not using the inaccessible true posterior, and because we are approximating the expectation with Monte Carlo samples from this approximation.
Note that the size of the variational parameter set $\psi$ is not included in the BIC calculation.
Note that we are using a positive log-likelihood convention, meaning a higher BIC is more favourable.

%
%

%% file: figures/CCN_figure3.tex
\begin{figure}[t!]
\begin{center}
  \includegraphics[width=\linewidth]{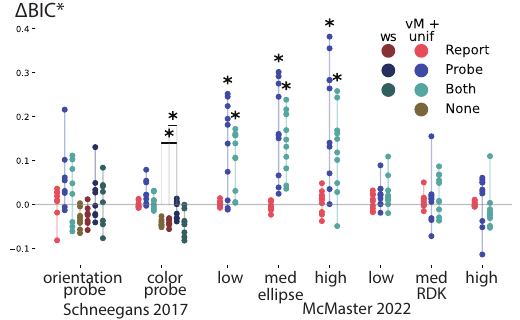}
\end{center}
\caption{
$\Delta\text{BIC}^*_s$ (see main text) on various datasets.
In all cases, the baseline model $\mathcal{M}'$ is a $\texttt{none}$ (flat) swap function with a von Mises emission and a uniform component (`vM. + unif'), and $\Delta\text{BIC}$ is compared for other swap function forms and for the wrapped stable (`ws') emission distribution.
$\ast$ indicates that the spread of this statistic is significantly ($p<0.05$) above 0, i.e. that the denoted model is a better fit to the data than the baseline model.
} 
\label{fig:full_model_comparison}
\end{figure}

%% file: 3_model_validation.tex
We start by validating the model fitting procedure described above by demonstrating the success of qualitative and quantitative parameter recovery when fitting to synthetic data.
We denote the parameters used to generate some synthetic data $\theta_\mathcal{M}$; the data it generates $\mathcal{D}[\theta_\mathcal{M}]$; and the parameters of the model fit to this data $\theta_{\mathcal{M}'}[\theta_\mathcal{M}]$
As before, $\mathcal{M},\mathcal{M}'\in\{\texttt{probe, report, both, none}\}$, refering to which feature dimensions the model architecture conditions swap errors on.
Again note that $\mathcal{M}=\texttt{none}$ is equivalent to the mixture model used by \citet{Bays2009} and \citet{Alleman2024}, and yields the same distribution over responses (equation \eqref{eq:actual_likelihood}) once component weights are marginalised out.

First, we show that models fit to data generated with swap functions conditioned on the same, or more, dimensions accurately recapitulate the shape of $f$ in those dimensions.
Figure \ref{fig:model_validation}A shows the true and recovered posterior approximations over one-dimensional swap functions when $\mathcal{M} = \mathcal{M}'$, for convex and concave dependence on either the probe or report dimension.
We see that BNS is able to accurately capture the ground truth generative process for a range of swap dependence modulation magnitudes and forms (with intuitive exceptions explained in the caption).
%
%
Conversely, we then show that model architectures fitted to data generated with swap functions conditioned on fewer, or different, feature dimensions naturally remove modulation of $f$ along dimensions of which the source data was independent.
Figure \ref{fig:model_validation}B shows that $\mathcal{M}=\texttt{probe},\mathcal{M}'=\texttt{report}$ (and vice versa) yields a flat swap function, effectively defaulting to $\texttt{none}$ as the fitted model cannot capture any of the variability in swap probability.
Figure \ref{fig:model_validation}C shows the converse - when $\mathcal{M}'=\texttt{both}$ is fitted to data from $\mathcal{M}=\texttt{probe}\}$, the fitted model removes most of the modulation from the redundant dimension.
In some cases, a synthetic dataset larger than the real dataset size (see \cite{McMaster2022} dataset in \textbf{Materials}) is required to achieve this effect completely.
This is a model limitation to which we return shortly.

Finally, we show that BIC is suitable for verifying the difference in fit quality for the latter scenario when training on a realistic number of trials, but can only verify the former for the $\texttt{both}$ 2D model in the limit of large synthetic datasets.
We define the metric $\Delta\text{BIC}_s(\mathcal{M}'; \mathcal{M})$ as the difference in average per-trial BIC for a `synthetic subject':
\begin{equation} 
\label{eq:delta_bic}
    \text{BIC}(\mathcal{D}_s[\theta_\mathcal{M}]; \theta_{\mathcal{M}'}[\theta_\mathcal{M}]) - 
    \text{BIC}(\mathcal{D}_s[\theta_\mathcal{M}];\theta_\mathcal{M}[\theta_\mathcal{M}])
\end{equation}
across multiple synthetic datasets $\mathcal{D}[\theta_\mathcal{M}]$.
This quantity summarises the ability of model $\mathcal{M}$ to capture the statistical structure of data generated by model $\mathcal{M}'$.
One would expect $\Delta\text{BIC}_s(\mathcal{M}'; \mathcal{M}) < 0$ for any $\mathcal{M}\neq\mathcal{M}'$, either because $\mathcal{M}'$ cannot condition swaps in the right dimensions, or because it penalised for its complexity.
Figure \ref{fig:model_validation_BIC} shows the distribution of $\text{BIC}_s(\mathcal{M};\mathcal{M}')$ across all combinations of model classes.
We see that a realistic amount of data per synthetic subject is sufficient to recover the model class (i.e. $\Delta\text{BIC} < 0$) in most cases, except for when the data is generated by $\mathcal{M}=\texttt{both}$.
In this case, it is necessary to move to large synthetic datasets when fitting $\mathcal{M}'$, with 10 times the number of trials, to reliably recover the generative model class.
This suggests that a classical BIC penalty term overly penalises the 2D model, motivating us to devise statistical tests which do away with a complexity penalty when fitting to real data.
We elaborate on this test, and present results, in the next section.

%% file: figures/CCN_figure4.tex
\begin{figure}[h!]
\begin{center}
  \includegraphics[width=0.9\linewidth]{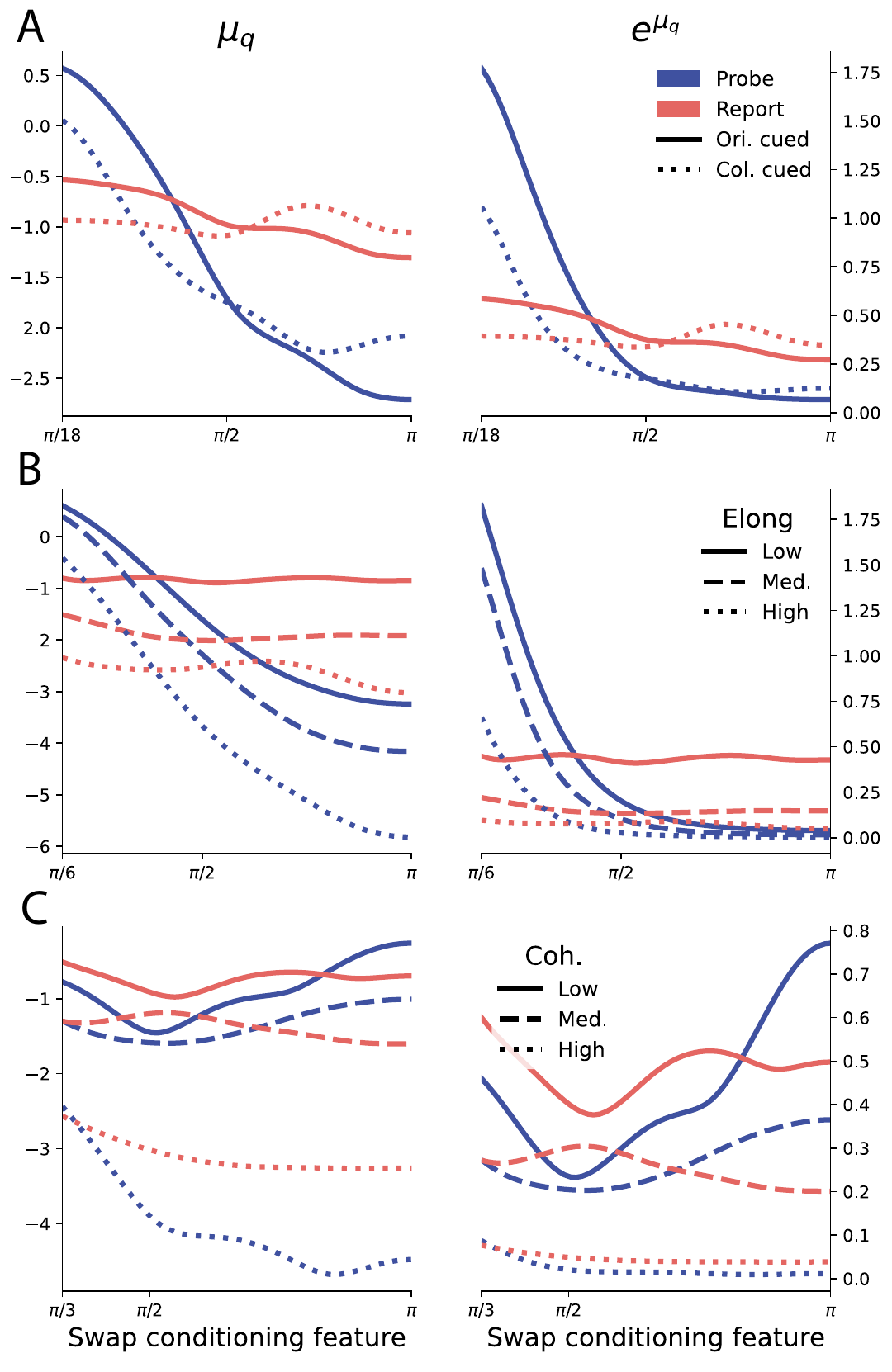}
\end{center}
\caption{
    Variational posterior mean functions $\mu_q=\mathbb{E}_q[f]$ for one dimensional models in the \textbf{A}: \cite{Schneegans2016} separated by role of each feature, and \textbf{B} orientation-cued and \textbf{C} direction-cued \cite{McMaster2022} datasets, separated by stimulus strengths (elongation and coherence respectively).
    We also show the exponent, to show how low evaluations of $f$ are compressed in the softmax function, but error intervals are omitted for clarity.
} 
\label{fig:visualise_real_data_fits}
\end{figure}

%% file: figures/CCN_figure5.tex
\begin{figure*}[h!]
\begin{center}
  \includegraphics[width=0.85\linewidth]{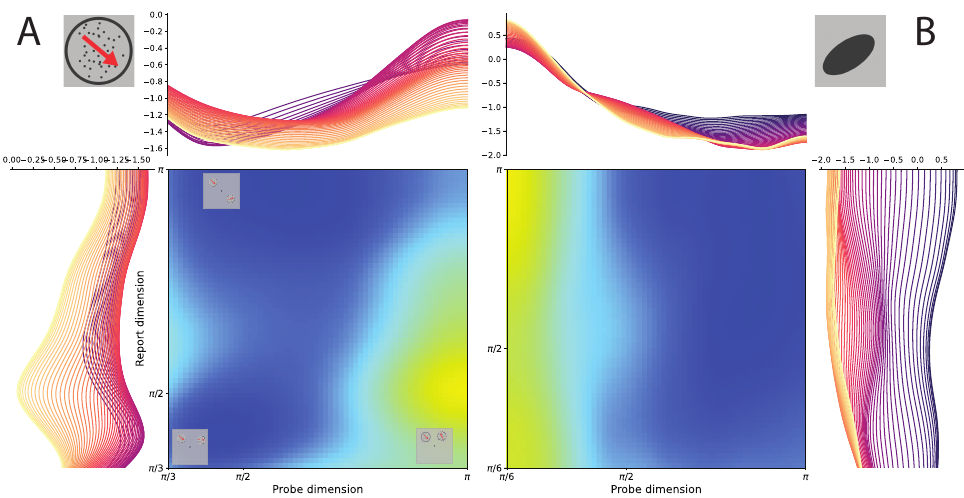}
\end{center}
\caption{
    Mean of $f$ when fit to \cite{McMaster2022} direction-cued RDK dataset (\textbf{A}) and orientation cued ellipse dataset (\textbf{B}) - location is recalled in both cases. 
    Consult Figure \ref{fig:model_validation} for a legend.
    Modulation in the report dimension, in particular a decay with location, indicates an encoding error mechanism of swaps for the RDK task.
    Example pairs of stimuli are shown at corresponding locations in $\boldsymbol{x}$-space to help visualise this (note that set-size is 4 for this task).
    All inset figures of stimuli are adapted from \cite{McMaster2022}.
} 
\label{fig:visualise_2D_real_data_fits}
\end{figure*}

%% file: 4a_datasets.tex
\subsubsection{Materials}
\label{sec:datasets}
In this section we show results when fitting BNS to real human behavioral data.
We again present both qualitative descriptions of the shape of the swap function approximate posterior $q(f)$ when fitted to data, and quantitative model comparison results across different swap dependency types, for multiple datasets.
We consider datasets from \citep{McMaster2022} (10 subjects, 2 experiments, 3 stimulus conditions, 96 trials/condition/subject/experiment), specifically datasets where location is reported, with ellipse orientation (set size = 6) or random dot kinematogram (RDK) motion direction (set size = 4) as the cue feature; and \cite{Schneegans2017} (8 subjects, 2 trial types, 60 trials/type/subject), where orientation is cued and colour is reported, or vice versa.
For the latter set of data, we consider both the von Mises and wrapped stable as emissions distributions, while for the former, given the high precision when reproducing location data, we limit residual distribution to the von Mises distribution.
The majority of our results are concordant with swap errors resulting from errors in cuing, manifested as a monotonic modulation of $f$ along the probe dimension.
However, when faced with a mixed dependence on both probe and report feature dimensions in one dataset, we turn to further result validation using synthetic data.

\subsubsection{Model comparison}
Figure \ref{fig:full_model_comparison} shows $\Delta\text{BIC}^*(\mathcal{M};\mathcal{M}')$:
\begin{equation}
    \text{BIC}(\mathcal{D}^*; \theta_\mathcal{M}[\mathcal{D}^*]) - 
    \text{BIC}(\mathcal{D}^*; \theta_{\mathcal{M}'}[\mathcal{D}^*])
\end{equation}
where $\mathcal{D}^*$ is the true data, and $\theta_\mathcal{M}[\mathcal{D}^*]$ are the parameters of model $\mathcal{M}$ when fit to the real data.
$\Delta\text{BIC}^*(\mathcal{M};\mathcal{M}')$ provides a direct comparison in the quality of fit of the two model types.
In Figure \ref{fig:full_model_comparison} we always take $\mathcal{M}'$ as the global baseline model - the case where swap errors are not conditioned on any feature values, and the emission distribution is von Mises.
As noted above, this collapses to a simple three-component model of responses \cite{Bays2009}.
For the orientation-cued \cite{McMaster2022} dataset, conditioning on the probe dimension consistently improves the quality of fit, and in the \cite{Schneegans2017} datasets there is a large, although not always significant, improvement within the wrapped stable family of model classes.
Because subject-specific emission parameters are more heavily penalised than shared swap function parameters, there is an asymmetry in comparing across emission distribution types.

%% file: 4b_simple_cue_recovery.tex
Comparing modulation in $\mathcal{M}=\texttt{probe, report}$ swap functions in Figure \ref{fig:visualise_real_data_fits}B (orientation-cued, location-recalled dataset) shows a clearcut example of cue similarity as the main or sole cause of swap errors.
The $\texttt{probe}$ model shows a strong monotonic increase in swap probability with proximity to the cue in the orientation dimension, which increases in magnitude as the probe dimension decreases in strength
The $\texttt{report}$ model removes all modulation in the mean of $q(f)$, akin to when it was fit to data explicitly generated by the $\texttt{probe}$ model in Figure \ref{fig:model_validation}B.
Figure \ref{fig:visualise_2D_real_data_fits}B shows an example 2D swap function, $\mathcal{M} = \texttt{both}$, for one of these datasets, bearing resemblance to Figure \ref{fig:model_validation}C, where the 2D model is fitted to data generated from conditioning on a single feature.
This effect is also seen in the two datasets from \cite{Schneegans2017} (Figure \ref{fig:visualise_real_data_fits}A), although there is minor modulation on colour when it is being recalled.
Importantly, the amplitude of $f$ for the $\texttt{probe}$ model varies greatly in the exponential space (relevant because the output of the swap function is subsequently normalised with a softmax) depending on which feature is probed, further implicating retrieval time as the source of swap errors.

%% file: 4c_non_trivialcue_recovery.tex
Figure \ref{fig:visualise_real_data_fits}C shows the $\texttt{probe}$ function mean for the RDK direction-cued, location-recalled dataset from \citep{McMaster2022}, revealing a strong non-monotonic dependence on cue dimension.
While initially appearing discordant with previous findings, this form of modulation follows from a cuing error in the event of erroneous encoding, maintenance in memory, or accessing of RDK directions as pure orientations.
BNS's ability to extract the form of this non-monotonic dependence, which combines cue proximity in the traditional sense (decreasing with angular distance) and directional stimuli erroneously remembered as orientations without directionality, is a product of its flexible, data-driven approach that imposes no prior constraints on how swap errors condition on distractor-cue relationships.
This non-monotonic pattern may already suggest that items are imperfectly encoded and maintained in memory from the outset, implicating these processes as additional mechanisms of swapping beyond just retrieval failures. 
We examine these implications more thoroughly in the following section.
%

%% file: figures/CCN_figure6.tex
\begin{figure*}[t!]
\begin{center}
  \includegraphics[width=0.95\linewidth]{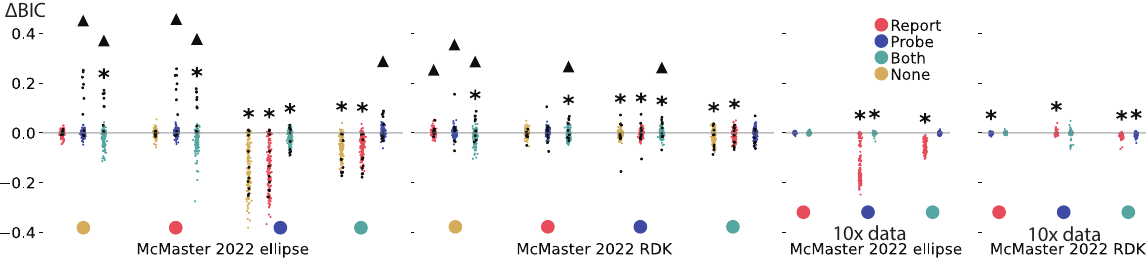}
\end{center}
\caption{
    Please consult Figure \ref{fig:model_validation_BIC} for interpreting the $\Delta \text{BIC}$ scattered here in colour.
    Here, 
    Black dots show $\Delta \text{BIC}^*_s$ (see main text) for the real subjects.
    $\blacktriangle$ indicates that the total $\Delta \text{BIC}^*(\mathcal{M};\mathcal{M}')$ is significantly ($p<0.05$) higher than the distribution of $\Delta \text{BIC}(\mathcal{M};\mathcal{M}')$ for different synthetic datasets, indicating the fitted model class $\mathcal{M}'$ better describes the data.
} 
\label{fig:synth_mcmaster_crossfits}
\end{figure*}

%% file: 5a_rdk_mixed_dependence.tex
Figure \ref{fig:visualise_real_data_fits}C shows a similar degree as modulation for $\mathcal{M}=\texttt{report}$ as for $\texttt{probe}$, when fitting to the direction-cued RDK dataset, e.g. at medium RDK coherence.
Fitting a $\mathcal{M}=\texttt{both}$ to one of these datasets shows that the two dependencies persist simultaneously (Figure \ref{fig:visualise_2D_real_data_fits}A).
In comparison to Figure \ref{fig:model_validation}C left ($\theta_\texttt{both}[\theta_\texttt{probe}]$), where, in fact, the data generating model was directly produced from training on the same data (i.e. $\theta_\texttt{probe} = \theta_\texttt{probe}[\mathcal{D}^*]$ - the limegreen line in Figure \ref{fig:model_validation}C left is the same as the medium coherence blue line in Figure), the modulation in the report dimension is much stronger and much more localised when fitting to the real data.
It also displays modulation at much higher absolute values, and across a much broader range of values, meaning the variability in logits it causes will survive the softmax operation.
However, the existence of any modulation when fitting to 1D synthetic data may casts doubt on the ability of BNS to prune out dependence in the report dimension when fitted to data with this specific report dimension swap error dependence without artifacts.
In the next section, we verify this dependence on the report dimension in the real data using further model validation, before going on to interpret this finding as evidence of swap errors caused before recall time.

%% file: 5b_rdk_mixed_dependence_validation.tex
Figure \ref{fig:synth_mcmaster_crossfits} shows $\Delta\text{BIC}_s$ for combinations of models where the generating model parameters were derived from data, that is $\theta_\mathcal{M} = \theta_\mathcal{M}[\mathcal{D}^*]$ in equation \ref{eq:delta_bic}.
As before, a significantly negative value indicates that the alternative model $\mathcal{M}'$ is not able to capture the variability in swap error dependence in $\mathcal{M}$, this time specifically for the dependence found in real data.
We consider two datasets here: orientation-cued, which BNS confirms exhibits pure probe-dependent swapping with no other factors, and direction-cued, which hints at mixed variability.
First, in the orientation-cued ellipse dataset, we see that variability in the $\texttt{report}$ swap function, like the simple model $\texttt{none}$, can be captured by any other model.
Similarly, variability learned by the 2D $\texttt{both}$ model can be fully explained by the $\texttt{probe}$ model, but not the $\texttt{report}$ model.
Even if we fit to a much larger amount of synthetic data, these properties of $\Delta\text{BIC}$ do not change for this dataset; the only change is that the relative complexity penalty difference decreases enough to remove the significant negativity of $\Delta\text{BIC}(\texttt{both}, \texttt{report})$.
This is unlike the scenario in Figure \ref{fig:model_validation_BIC}, where the parameters of the generating $\texttt{both}$ model were chosen to have swap function modulation along both dimensions.
Overall, this indicates that swap variability in this dataset is entirely described by the probe dimension.
In the realistic data regime, the same is true for the direction-cued RDK dataset.
To further examine this case in the light of a potentially missuited BIC penalty (Figure \ref{fig:model_validation_BIC}), we devise another statistical test for $\Delta\text{BIC}^*$ beyond its deviation from zero.
Calculated across multiple synthetic datasets, $\Delta\text{BIC}(\mathcal{M}';\mathcal{M})$ provides a null distribution for $\Delta\text{BIC}^*(\mathcal{M}';\mathcal{M})$, the same value but evaluated on real data.
A significantly high value of $\Delta\text{BIC}^*$ (shown in black in Figure \ref{fig:synth_mcmaster_crossfits}) indicates that the relative advantage of the fitted model $\mathcal{M}'$ on the real data is higher than that on synthetic data generated by $\mathcal{M}$, in turn suggesting that the synthetic data generated by $\mathcal{M}$ fails to recreate the statistical structure present in the real data, which $\mathcal{M}'$ is better equipped to explain.
Importantly, this test removes the bias of the BIC penalty, which we previous established to be unsuitable for comparing the 1D and 2D model classes.
This test differentiates the two datasets.
In the orientation-cued dataset, the null hypothesis is rejected when $\mathcal{M}'$ conditions on the probe dimension.
In the direction-cued dataset, increasing the dimensionality of $f$ strictly improves the statistical structure replicated by synthetic data, according to this statistical test.
When unhindered by complexity penalty, conditioning on both features better describes the statistical structure of the RDK dataset.
Finally, Figure \ref{fig:synth_mcmaster_crossfits} right shows the large synthetic dataset case for the direction-cued dataset.
Even in the large data limit, a one-dimensional swap function fails to capture the variability in the other dimension - either by itself or as part of a $\texttt{both}$ model.
This is further evidence that the modulation along the report dimension captures real statistical structure in the direction-cued dataset.

%% file: 6a_significance.tex
We provide a data-driven alternative to previous neural models \citep{Schneegans2017, McMaster2022}, free of assumptions about the parametric form of swap error dependence on distractors.
Fit to data, BNS first discovered a non-monotonic dependence of swap errors on cue distance for RDK stimuli which may erroneously be stored as pure orientations.
This was done automatically, without a metric distance defined on the circle of RDK motion directions, which is non-trivial for this form of stimulus, but would be a necessary preliminary for existing models.
%
BNS was then used to find a dual dependence of swap errors on both cue feature (RDK motion direction) distance and report feature (location) distance.
Specifically, it found that swaps were most likely when a distactor is near to the cued item in space, but with motion in the opposite direction.
For a location recall task, this implies swap errors in this dataset were made due to an error in encoding at stimulus presentation time.
This provides an alternative mechanism for swapping besides a retrieval error, which BNS was also able to recapitulate in other datasets.
The form of this function, which is not factorisable between the two feature dimensions, further shows that there is an interaction between the effects of probe proximity and report proximity, which again BNS has automatically captured.

%% file: 6b_limitations.tex
BNS is data-hungry, forcing us to pool the swap function architecture across subjects.
This is suboptimal, and ideally, we would fit a separate variational approximation $q_s(f)$ to each subject.
In practice, this has lead to underfitting (or over-regularisation) and deference of $q_s$ to the GP prior, due to the lack of data.
Preliminary investigations showed that the same was true for a hierarchical model, where a shared swap function $f$ defines a prior for subject-specific functions $f_s$ (e.g. $f$ has mean zero, and each $f_s$ has mean $f$).

BNS's future contribution to modeling psychophysics is two-fold.
First, it can be used to drive experimental design.
Based on our findings, we suggest further targeted investigation into the role of encoding error is necessary.
Second, BNS can be used to constrain and train neural models.
Trained neural network models are now commonplace in the neuroscience literature, superseding hand-tuned neural networks in their flexibility to capture neural dynamics \cite{Stroud2023}, including for modeling biases and errors in working memory and cognition tasks and predicting their neural mechanisms \citep{Cueva2021, MolanoMazn2023, Xie2023}.
By summarising the statistical structure in the data without assumptions on neural mechanisms behind the behavior, data-driven probabilistic models can be used to provide training data for such network models.